\documentclass[fleqn,usenatbib]{mnras}

\usepackage{newtxtext,newtxmath}

\usepackage[T1]{fontenc}

\DeclareRobustCommand{\VAN}[3]{#2}
\let\VANthebibliography\thebibliography
\def\thebibliography{\DeclareRobustCommand{\VAN}[3]{##3}\VANthebibliography}

\usepackage{graphicx}	
\usepackage{amsmath}	
\usepackage{subcaption}
\usepackage{dsfont}
\usepackage{booktabs}
\usepackage{orcidlink}
\usepackage[colorinlistoftodos, color=lime]{todonotes}
\DeclareMathAlphabet{\mathcal}{OMS}{cmsy}{m}{n} 

\title[Mitigating Time-Varying Systematics]{A Bayesian Method to Mitigate the Effects of Unmodelled Time-Varying Systematics for 21-cm Cosmology Experiments}

\author[C. J. Kirkham et al.]{
Christian J. Kirkham \textsuperscript{\orcidlink{0000-0001-5385-6329}},$^{1,2}$\thanks{E-mail: \href{mailto:cjk55@cam.ac.uk}{cjk55@cam.ac.uk}} 
Dominic J. Anstey \textsuperscript{\orcidlink{0000-0003-1742-7417}},$^{1,2}$\thanks{E-mail: \href{mailto:da401@cam.ac.uk}{da401@cam.ac.uk}} and Eloy de Lera Acedo \textsuperscript{\orcidlink{0000-0001-8530-6989}}$^{1,2}$\thanks{E-mail: \href{mailto:ed330cam.ac.uk}{ed330@cam.ac.uk}}
\\
$^{1}$Astrophysics Group, Cavendish Laboratory, J. J. Thomson Avenue, Cambridge, CB3 0HE, UK\\
$^{2}$Kavli Institute for Cosmology, Madingley Road, Cambridge, CB3 0HA, UK\\
}

\date{Accepted XXX. Received YYY; in original form ZZZ}

\pubyear{2023}

\begin{document}
\label{firstpage}
\pagerange{\pageref{firstpage}--\pageref{lastpage}}
\maketitle

\begin{abstract}
Radio observations of the neutral hydrogen signal from the Cosmic Dawn and Epoch of Reionisation have helped to provide constraints on the properties of the first stars and galaxies. Since this global 21-cm cosmological signal from the Cosmic Dawn is effectively constant on observing timescales and since effects resulting from systematics will vary with time, the effects of these systematics can be mitigated without the need for a model of the systematic. We present a method to account for unmodelled time-varying systematics in 21-cm radio cosmology experiments using a squared-exponential Gaussian process kernel to account for correlations between time bins in a fully Bayesian way. We find by varying the model parameters of a simulated systematic that the Gaussian process method improves our ability to recover the signal parameters by widening the posterior in the presence of a systematic and reducing the bias in the mean fit parameters. When varying the amplitude of a model sinusoidal systematic between 0.25 and 2.00 times the 21-cm signal amplitude and the period between 0.5 and 4.0 times the signal width, we find on average a 5\% improvement in the root mean squared error of the fitted signal. We can use the fitted Gaussian process hyperparameters to identify the presence of a systematic in the data, demonstrating the method's utility as a diagnostic tool. Furthermore, we can use Gaussian process regression to calculate a mean fit to the residuals over time, providing a basis for producing a model of the time-varying systematic.
\end{abstract}

\begin{keywords}
methods: data analysis -- cosmology: dark ages, reionization, first stars -- cosmology: early Universe
\end{keywords}



\section{Introduction}

One of the most promising probes of the physics of the Comsic Dark Ages, Cosmic Dawn and the Epoch of Reionisation is 21-cm cosmology \citep{furlanettoCosmologyLowFrequencies2006}. Approximately 379,000 years after the Big Bang the universe underwent a phase change known as `recombination' whereby the ionised electrons and protons combined to fill the universe with neutral hydrogen (HI), the redshifted afterglow of which can be observed as the Cosmic Microwave Background (CMB) radiation \citep{planckcollaborationPlanck2018Results2020a}. This HI gas has a hyperfine transition which emits and absorbs radiation at a wavelength of $\lambda=21$ cm or a frequency of $\nu = 1420$ MHz in the rest frame. We can define a statistical `spin temperature' which is related to the occupancy of the excited and neutral states of the HI gas. This signal is measured relative to the CMB temperature and is either in absorption or emission based on the coupling between the gas temperature, the background radiation and the spin temperature \citep{wouthuysenExcitationMechanism21cm1952,fieldSpinTemperatureIntergalactic1959, furlanettoCosmologyLowFrequencies2006}.

There are several low-frequency radio experiments which are attempting to detect the cosmological 21-cm signal. Interferometers such as HERA \citep{deboerHydrogenEpochReionization2017}, LOFAR \citep{haarlemLOFARLOwFrequencyARray2013}, the MWA \citep{tingayMurchisonWidefieldArray2013} and the the future SKA Observatory \citep{dewdneySquareKilometreArray2009} use arrays of telescopes to measure the spatial power spectrum of fluctations in the early universe. Experiments which measure the global sky-averaged 21-cm signal such as EDGES \citep{bowmanEmpiricalConstraintsGlobal2008}, SARAS \citep{singhSARASSpectralRadiometer2018,singhDetectionCosmicDawn2022}, LEDA \citep{priceDesignCharacterizationLargeaperture2018}, PRIZM \citep{philipProbingRadioIntensity2019}, MIST \citep{monsalveMapperIGMSpin2023} and REACH \citep{deleraacedoREACHRadiometerDetecting2022a} are working to place constraints on the physics of the Cosmic Dawn and the Epoch of Reionisation.

The only detection of the cosmological global 21-cm signal claimed so far is by the EDGES experiment \citep{bowmanAbsorptionProfileCentred2018}. This experiment is made up of two low-band dipole antennae located in the Murchison Radio Astronomy Observatory (MRO) in Western Australia which operate between 50 and 100 MHz \citep{bowmanEmpiricalConstraintsGlobal2008}. Since the detected signal had an abnormally flat and deep profile -- at least two times deeper than previous predictions \citep{reisSubtletyLyPhotons2021a} -- concerns were raised regarding the validity of the analysis methods and the effect of systematics \citep{hillsConcernsModellingEDGES2018, singhRedshifted21Cm2019, simsTestingCalibrationSystematics2020}. Another 21-cm global signal experiment, SARAS3 \citep{nambissanSARASCDEoR2021}, recently placed constraints on the 21-cm signal, rejecting the EDGES detection with 95.3\% confidence \citep{singhDetectionCosmicDawn2022}.

\cite{hillsConcernsModellingEDGES2018} found issue with the foreground modelling method used by the EDGES team. By comparing the EDGES foreground model with a physically motivated non-linear expression, they found that the optical depth of the ionosphere and the electron temperature are both negative indicating that the foreground fit is unphysical. They suggest that these unphysical values result from unaccounted for systematics in the data. Removing a 12.5 MHz sine wave from the data allows good fit to a broad Gaussian absorption profile, obtained with five foreground parameters. It is proposed that a sinusoid in the data -- which can be explained by any number of instrumental systematics (see Section \ref{s:REACH_systematics}) -- is what resulted in the \cite{bowmanAbsorptionProfileCentred2018} absorption profile with a flattened bottom which is consistent with the results of \cite{simsTestingCalibrationSystematics2020}.

Others have interpreted the EDGES signal as a need to introduce new exotic physics, as current theories cannot explain why there would be such a large contrast between the CMB temperature and the gas kinetic temperature. One such theory proposes dark matter of which a small fraction is millicharged e.g. \cite{barkanaPossibleInteractionBaryons2018}, which would scatter of the baryonic matter, providing an additional cooling mechanism. An alternative theory does not invoke new physics but rather suggests that there is an unaccounted for radio background \citep{fengEnhancedGlobalSignal2018}. This suggestion may be supported by measurements by ARCADE-2 and LWA \citep{fixsenARCADEMEASUREMENTABSOLUTE2011, dowellRadioBackground1002018}.

In this work we will investigate the effects of time-varying systematics and test a method to help identify systematics and mitigate their effects using Gaussian processes. In particular, we will perform this investigation in the context of the REACH global experiment \citep{deleraacedoREACHRadiometerDetecting2022a}. In Section \ref{s:methods} we will present the standard REACH pipeline and likelihood, introduce the Gaussian process likelihood and demonstrate how they could be used to perform time regression. In Section \ref{s:results} we present the results of introducing simulated systematics to the data and compare the signal recovery of both the standard and Gaussian process methods. In Section \ref{s:conclusions} we present the conclusions of the investigation.

\subsection{Systematics in the REACH System} \label{s:REACH_systematics}

\begin{figure*}
	\centering
	\includegraphics[width=0.9\linewidth]{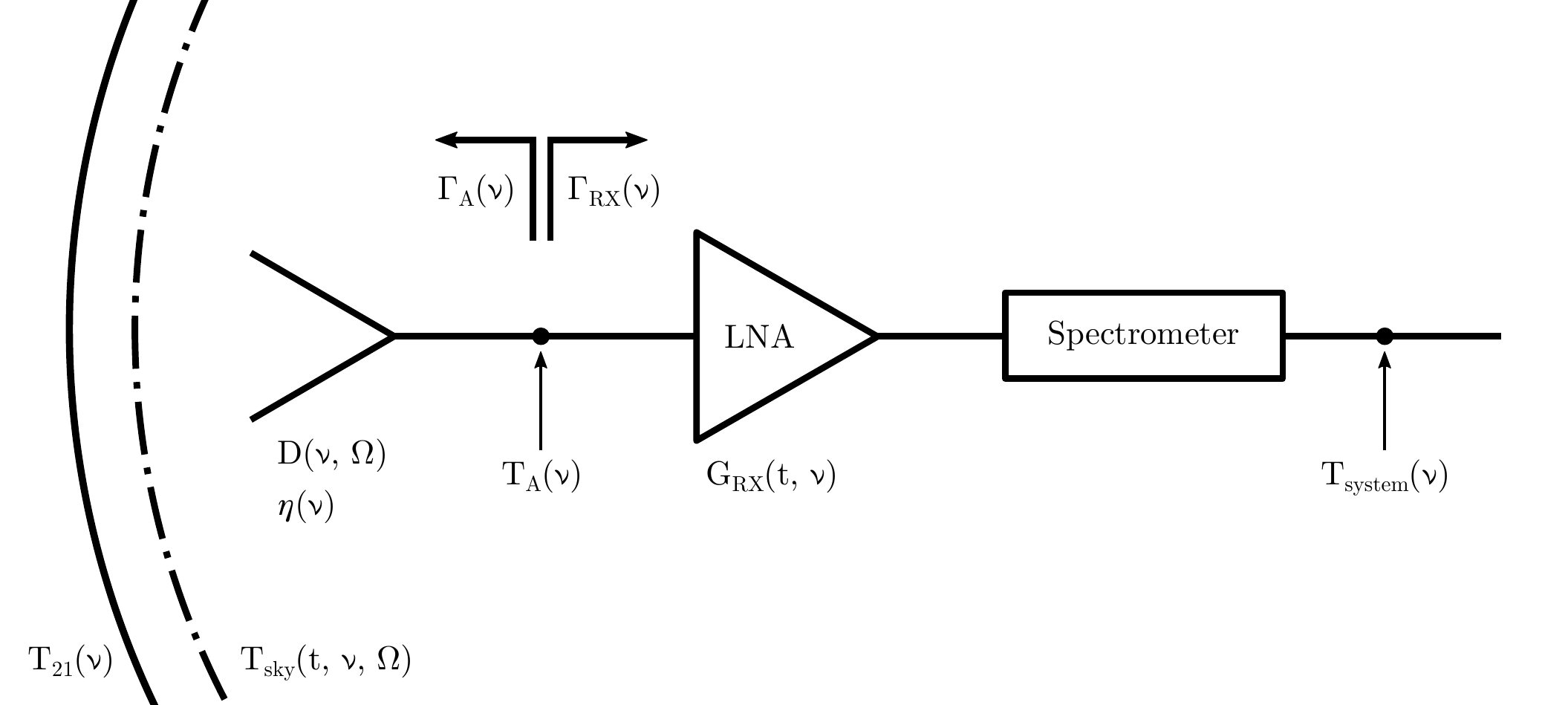}
	\caption{A schematic of the REACH antenna and receiver. Adapted from \protect\cite{cumnerRadioAntennaDesign2022}.}
	\label{f:antenna_systematics}
\end{figure*}

While all best efforts have been made to calibrate the REACH instrument \citep{roqueBayesianNoiseWave2021, ansteyGeneralBayesianFramework2021a, ansteyUseTimeDependent2022a, bevinsMAXSMOOTHRapidMaximally2021, scheutwinkelBayesianEvidencedrivenDiagnosis2022a, razavi-ghodsReceiverDesignREACH2023}, it might be inevitable that some systematics will end up in the final data so it is important that we understand and attempt to mitigate the effects of these unknown systematics. As was indicated in \cite{hillsConcernsModellingEDGES2018} and \cite{simsTestingCalibrationSystematics2020}, not accounting for systematics can potentially have a large impact on your final fit. In particular for this investigation we will have to consider the effects of the galactic foreground moving over the sky and the beam, and the temperature changing over time as both will potentially introduce systematics in the data that will vary with time.

Figure \ref{f:antenna_systematics} shows a schematic of the REACH antenna and receiver system. Here, $D(\nu, \Omega)$ refers to the directivity of the antenna, $\Gamma_\text{A} (t, \nu)$ is the impedence of the antenna, $\Gamma_\text{RX} (t, \nu)$ is the impedence of the receiver, $G_\text{RX} (t,\nu)$ is the gain of the Low Noise Amplifier (LNA), $\eta(t, \nu)$ is the radiation efficiency of the antenna, $T_\text{A} (\nu)$ is the antenna temperature and $T_\text{system} (\nu)$ refers to system temperature that is produced by the receiver. These components combine to make the time-dependent antenna temperature \citep{cumnerRadioAntennaDesign2022},
\begin{equation}
    T_\text{ant} (\nu, t) = \frac{1}{4\pi} \int_\Omega D(\nu, \Omega) \eta(\nu)(T_\text{sky} (t, \nu, \Omega) + T_{21} (\nu)) \phantom\cdot\mathrm{d}\Omega
\end{equation}
which can be integrated to give the time-averaged antenna temperature,
\begin{equation}
    T_\text{A} = \int_t T_\text{ant} (\nu, t) \phantom\cdot \mathrm{d} t.
\end{equation}
Including impedence reflections and the noise term, $N(t, \nu, \Gamma_\text{A})$, gives the system temperature,
\begin{equation}
    T_\text{system} (\nu) = \int_t (T_\text{ant}  (1 - |\Gamma_\text{A}|^2) G_\text{RX} (t, \nu) + N(t, \nu, \Gamma_\text{A})) \phantom\cdot \mathrm{d}t.
\end{equation}

The $1 - |\Gamma_\text{A}|^2$ term arises from unmatched impedence between the antenna and the receiver. Each cable of the antenna has its own impedence value and if there is a difference in impedence at the interface between the two cables or devices then electrical signals will be partially reflected off this interface. As a result, a standing wave may form in the cables producing a sinusoidal systematic in the data. 

Cable reflections can also produce sinusoidal systematics from noise sources such as the LNA noise or sky temperature noise. This is problematic since systematics will remain in the averaged data while noise can usually be integrated down, with the amplitude of the noise scaling as $1/\sqrt{t_\text{int}}$ where $t_\text{int}$ is the integration time of the instrument \citep{krausRadioAstronomyReceivers1986}. Similarly, radio-frequency interference (RFI) -- which can usually be flagged \citep{leeneyBayesianApproachRFI2022} and the relevant frequency bin excised -- can be made sinusoidal when it passes through a system with unmatched impedences.

\begin{figure}
     \centering
     \includegraphics[width=\linewidth]{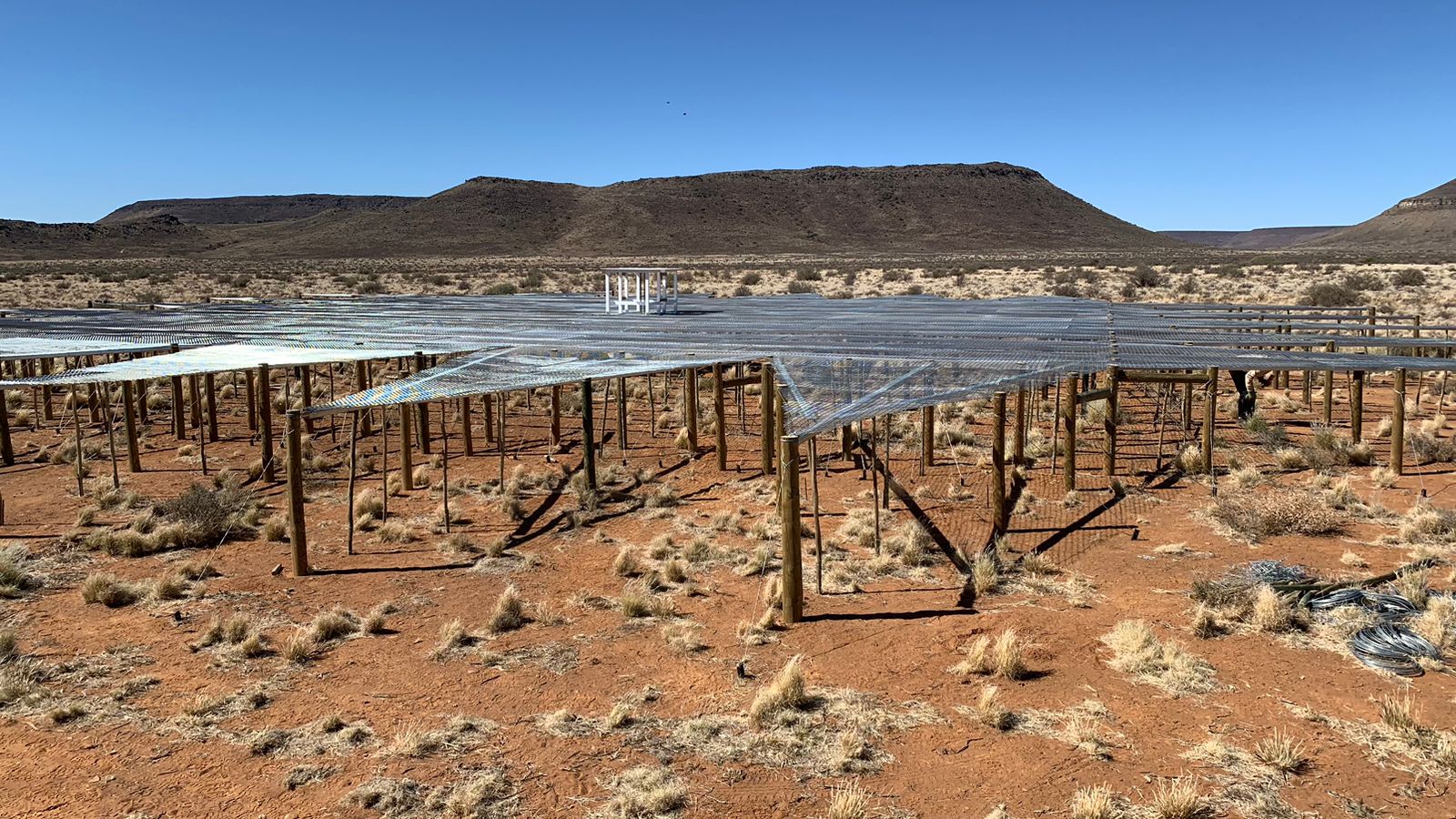}
    \caption{The REACH antenna and ground plane in the Karoo desert in South Africa. The size of the ground plane is 25m by 25m with a 5m serrations and is built 1m above the ground.}
    \label{f:reach_photo}
\end{figure}

Other systematics can arise from incorrectly modelling the directivity pattern of the antenna or the sky temperature, the residuals of which could be approximately sinusoidal in form \citep{priceDesignCharacterizationLargeaperture2018}. The REACH pipeline uses the foreground fitting procedure to correct for this somewhat, although limitations -- such as the antenna having such a high chromaticity that it exceeds the corrective abilities of the foreground fitting -- mean that this cannot be done perfectly accurately \citep{ansteyGeneralBayesianFramework2021a}. Reflections from the soil due to its dielectric properties can also result in sinusoidal systematics as standing waves form between the antenna and the ground \citep{bevinsMAXSMOOTHRapidMaximally2021, bevinsComprehensiveBayesianReanalysis2022}. This is particularly problematic as the exact dielectric constant of the soil is unknown and, as such is difficult to be modelled accurately \citep{pattisonModellingHotHorizon2023}. Soil reflections are somewhat mitigated by the inclusion of a 25 m by 25 m square ground plane underneath the antenna, pictured in figure \ref{f:reach_photo}, although finite ground planes can introduce new standing waves \citep{bolliPreliminaryAnalysisEffects2020}.

Of particular interest to this investigation is the question of what happens if the systematic changes with time. An example of this effect was found by the LEDA team who discovered that the pattern of oscillations changed after rainfall \citep{priceDesignCharacterizationLargeaperture2018}, potentially resulting from the soil's moisture content changing the dielectric properties of the ground. Changing systematics could also arise from cable reflections which reflect the foreground power and, as such, results in a sinusoidal systematic whose amplitude is modulated over time as the Earth rotates. As impedence depends on temperature and components may flex as they warm or cool, the environmental conditions can also affect systematics introduced by the receiver \citep{cumnerRadioAntennaDesign2022}.

\section{Methods} \label{s:methods}

\subsection{Bayesian Inference}\label{s:intro_bayes}

Bayesian inference is a statistical method which can be used to infer the probability distribution of an unknown variable from some given dataset and as such is a useful tool for parameter estimation. The method relies on applying Bayes' theorem for inverting conditional probabilities which can be expressed as
\begin{equation}
    P(\mathbf \theta|\mathbf D, \mathcal M) = \frac{P(\mathbf D|\mathbf \theta, \mathcal M) \cdot P(\mathbf \theta | \mathcal M)}{P(\mathbf D | \mathcal M)},
\end{equation}
where $\mathbf \theta$ are the parameters of the model, $\mathcal M$, that we're trying to fit and $\mathbf D$ is the vector of data points \citep{siviaDataAnalysisBayesian2006}. $P(\mathbf \theta | \mathcal M)$, or $\pi (\mathbf \theta)$, is known as the `prior distribution' and represents our prior knowledge of the parameter probability distribution. $P(\mathbf D|\mathbf\theta, \mathcal M)$, or $\mathcal L (\mathbf \theta)$, is known as the `likelihood' and represents the probability of observing the dataset given the chosen model and parameters are true. $P(\mathbf\theta|\mathbf D, \mathcal M)$, or $\mathcal P (\mathbf \theta)$, is known as the `posterior distribution' and is the probability of the parameters given the data and the model, and is inferred from the prior and likelihood distributions. Finally, $P(\mathbf D|\mathcal M)$ is called the `Bayesian evidence', sometimes given as $\mathcal Z$, and can be used as a goodness-of-fit measure for model comparison.

The likelihood function is an expression of how likely the data is given the model and its form depends on the probability distribution of the data. If the data is randomly distributed according to a multivariate Gaussian distribution, we use a Gaussian likelihood function of the form
\begin{equation} \label{e:Gaussian_likelihood}
    \mathcal L(\mathbf \theta) = \frac{1}{\sqrt{(2\pi)^n|\mathbf C|}}\exp\left(-\frac{1}{2}(\mathbf D - \mathbf M(\mathbf \theta))^T\boldsymbol{C}^{-1}(\mathbf D - \mathbf M(\mathbf \theta))\right),
\end{equation}
where $\mathbf M(\mathbf \theta)$ is the model function, $n$ is the length of the data, $\mathbf D$ and $\mathbf C$ is the covariance matrix.

The Bayesian evidence can then be calculated by integrating over the parameter space, a technique known as `marginalising', as
\begin{equation} \label{e:evidence_integral}
    \mathcal Z = P(\mathbf D|\mathcal M) = \int \mathcal L(\mathbf \theta) \cdot \pi (\mathbf \theta) \phantom\cdot \mathrm{d} \mathbf \theta.
\end{equation}
To calculate the evidence and sample the posterior we use the nested sampler \textsc{PolyChord} \citep{handleyPolychordNestedSampling2015, handleyPolychordNextgenerationNested2015}.

\subsection{REACH Pipeline}

 The REACH pipeline uses a framework for jointly modelling galactic foregrounds and correcting for chromaticity \citep{ansteyGeneralBayesianFramework2021a}. As the hexagonal dipole is not an achromatic beam, the antenna has a directivity pattern, $D(\Omega, \nu)$ which depends on the direction of the observation, $\Omega$ and the radio frequency, $\nu$. This is then convolved with the time-dependent sky temperature, $T_\text{sky} (\Omega, \nu, t)$, at time of observation, $t$, to get the observed antenna temperature,
\begin{equation}
    T_\text{data} (\nu) = \frac{1}{4\pi} \int_0^{4\pi} D(\Omega, \nu) \int^{t_\text{end}}_{t_\text{start}} T_\text{sky} (\Omega, \nu, t)\phantom\cdot \mathrm d t \mathrm d \Omega + \hat \sigma,
\end{equation}
where $\hat \sigma$ is noise, assumed here to be uncorrelated Gaussian noise. The observed sky temperature model used in the pipeline is made up three main components,
\begin{equation}
    T_\text{sky} (\Omega, \nu, t) = T_\text{fg} (\Omega, \nu, t) + T_\text{sg} (\nu) + T_\text{CMB},
\end{equation}
where $T_\text{fg} (\Omega, \nu, t)$ is the galactic foreground temperature, $T_\text{sg} (\nu)$ is the temperature of the global 21-cm signal and $T_\text{CMB} = 2.73$ K is the CMB temperature. 

Due to synchrotron radiation emitted by hot gas in the galaxy, the galactic foreground emission must be modelled as it is $\sim10^4$ times larger in magnitude than the 21-cm signal \citep{shaverCanReionizationEpoch1999}. Since there are no foreground emission maps in the REACH band, the pipeline uses a global sky map (GSM) of antenna temperature at 230 MHz \citep{deoliveira-costaModelDiffuseGalactic2008}. While the full REACH pipeline decomposes the sky into regions of uniform spectral index, to reduce computational time we simulate the sky as having a single spectral index, $\beta = -2.55$. The resulting sky map in the REACH band is then found as
\begin{equation}
    T_\text{fg} (\Omega, \nu) = (T_{230} (\Omega) - T_\text{CMB}) \left(\frac{\nu}{230\text{ MHz}}\right)^{-\beta},
\end{equation}
where $\beta$ is fitted for as free parameter. We have defined the map using the 230 MHz GSM, although the 408 MHz map is an equally appropriate choice and produces very little difference in results. 

The Bayesian evidence and posterior samples of the foreground and signal models are found using \textsc{PolyChord}, with uniform priors in the range $2.45844 < \beta < 3.14556$, the full range of a spectral index map derived using the 230 MHz and 408 MHz GSMs \citep{deoliveira-costaModelDiffuseGalactic2008,ansteyGeneralBayesianFramework2021a}. Testing of the pipeline is done using simulated data which is calculated using the 230 MHz GSM and the spectral index map. A Gaussian mock 21-cm global signal of the form
\begin{equation} \label{e:Gaussian_signal}
    T_\text{sg} (\nu) = -A_{21} \exp\left( -\frac{(\nu - \nu_c)^2}{2\sigma_{21}^2} \right),
\end{equation}
where $A_{21}$ is the signal amplitude, $\nu_c$, the centre frequency and $\sigma_{21}$, the signal width, was added to the simulated data. As this is a similar shape to the physical 21-cm signal, this is a suitable analogue for testing purposes and is vastly less computationally expensive to model than other more physically motivated models. The signal parameters have uniform priors in the ranges $50 < \nu_c < 200$ MHz, $10 < \sigma_{21} < 20$ MHz and $0 < A_{21} < 0.25$ K. When testing the pipelines, they will be run with a 21cm signal with parameters $A_{21} = 0.155$ K, $\sigma_{21} = 15$ MHz and $\nu_c = 80$ MHz. Gaussian noise with standard deviation $\hat \sigma = 0.1$ K is added to the data to simulate the uncorrelated noise of the system.

The REACH pipeline's method of fitting of the foreground parameter can make use of time dependent data in the pipeline \citep{ansteyUseTimeDependent2022a}. The observation is split into $N_t$ consecutive integrations, or time bins, which are measured in local sidereal time (LST) to match the observation to the stage of Earth's rotation. We hence define the likelihood like so,
\begin{align} \label{e:standard_pipeline_likelihood}
\begin{split}
    \log \mathcal L_\text{std} &= \sum_i \sum_j -\frac{1}{2} \log(2\pi\sigma_{0,\text{std}}^2)
    \\&-\frac{1}{2} \left( \frac{T_\text{data}(\nu_i, t_j) - (T_\text{fg} (\nu_i, t_j) + T_{21} (\nu_i) + T_\text{CMB})}{\sigma_{0,\text{std}}} \right)^2,
\end{split}
\end{align}
where $i$ refers to the $i$th frequency bin, and $j$ to the $j$th time bin. This likelihood will be referred as the `standard pipeline' from here on.

\subsection{Systematic Model}

As discussed in section \ref{s:REACH_systematics}, we might expect to find systematics in the REACH system which are sinusoidal, generated by standing waves in the receiver or on the antenna. As a result we can model the general systematic we may expect to see as a damped sinusoid of the form
\begin{equation} \label{e:systematic_model}
    T_\text{sys} (\nu) = A_\text{sys} \left(\frac{\nu}{\nu_{0,\text{sys}}}\right)^{-\alpha_\text{sys}} \sin \left(\frac{2\pi \nu}{P_\text{sys}} + \phi_\text{sys}\right),
\end{equation}
where $\nu_{0,\text{sys}} = 50$ MHz is the fiducial radio frequency of the systematic, $A_\text{sys}$ is the amplitude of the systematic, $P_\text{sys}$ is the period of the systematic, $\phi_\text{sys}$ is the phase of the systematic, and $\alpha_\text{sys}$ is the dampening of the systematic \citep{scheutwinkelBayesianEvidencedrivenDiagnosis2022a}. In this paper the value of the dampening is fixed at $\alpha_\text{sys} = 1.4$.

The model of the time-varying systematic we will consider in this paper is the case where a systematic is constant in phase, frequency and dampening but modulates its amplitude according to the incoming power from the galactic foreground. Here we define the amplitude of the systematic at time bin $j$ as 
\begin{equation} \label{e:fg_modulated_systematic}
    A_\text{sys} (t_j) = A_\text{sys} (t_0) \cdot \frac{T_\text{fg} (\nu = \nu_0, t = t_j)}{T_\text{fg} (\nu = \nu_0, t = t_0)},
\end{equation}
where $\nu_0$ determines the radio frequency from which the foreground power is taken. Here, we will take $\nu_0 = 50$ MHz. Currently the modulation is done monochromatically although it may be better to model the systematic by modulating each frequency bin separately in future.

\begin{figure}
     \centering
     \includegraphics[width=\linewidth]{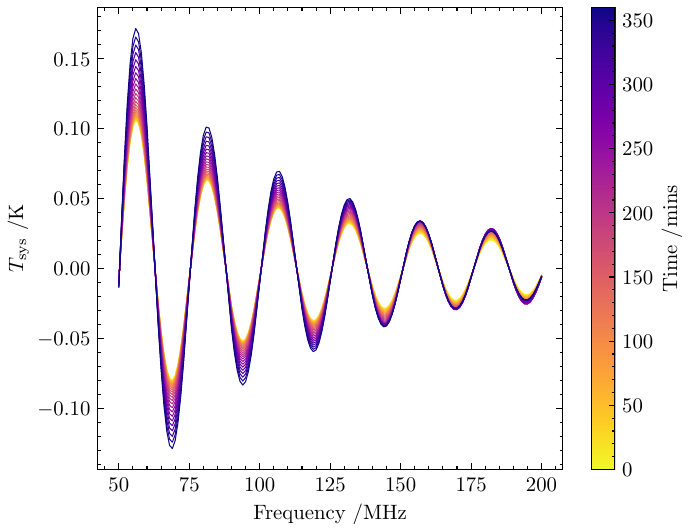}
    \caption{Example of a damped sinusoidal systematic given by equations \ref{e:systematic_model} and \ref{e:fg_modulated_systematic} which is modulated by the incoming power from the galactic foreground over 24 time bins of length 15 minutes. The initial amplitude of the systematic was set to $A_\text{sys} = 0.209$ K.}
    \label{f:sys_fg_mod_avg_example}
\end{figure}

Figure \ref{f:sys_fg_mod_avg_example} shows this foreground-modulated systematic for 24 time bins of length 15 minutes. Here, the systematic shows a slight shift in amplitude as the foreground power increases over time. As there is no change in the phase or frequency of the systematic over time, the signal does not average down by any significant amount and as such its amplitude cannot be affected by increasing integration times.

\subsection{Gaussian Processes}

In order to account for the covariance in the model residuals introduced by the presence of a systematic in the data, we will use Gaussian processes (GPs) to build upon the standard REACH likelihood. Gaussian processes are non-parametric probabilistic methods of performing regression and forecasting that have found particular use for Bayesian time series regression \citep{mackayInformationTheoryInference2002,robertsGaussianProcessesTimeseries2013}. 

We define a GP as a collection of random variables which have consistent joint Gaussian distributions \citep{rasmussenGaussianProcessesMachine2004}. These Gaussian distributions are defined by the covariance function, or kernel, of the Gaussian process where the choice of kernel is arbitrary, depending on the data you are trying to model. There is a wealth of literature into the variety of structures of kernel that can be used for GP regression, where their applicability depends on the problem that is trying to be solved.

The kernel, which we will use in this paper is the squared exponential,
\begin{equation} \label{e:SE_kernel}
    K_\text{SE}(t_i,t_j) = \sigma_\text{SE}^2 \exp \left( -\frac{|t_i - t_j|^2}{2\ell^2} \right),
\end{equation}
where $\ell$ is known as the characteristic length scale of the Gaussian process and $\sigma_\text{SE}^2$ is the scale factor of the squared exponential kernel \citep{rasmussenGaussianProcessesMachine2006}. 

\begin{figure}
     \centering
     \includegraphics[width=\linewidth]{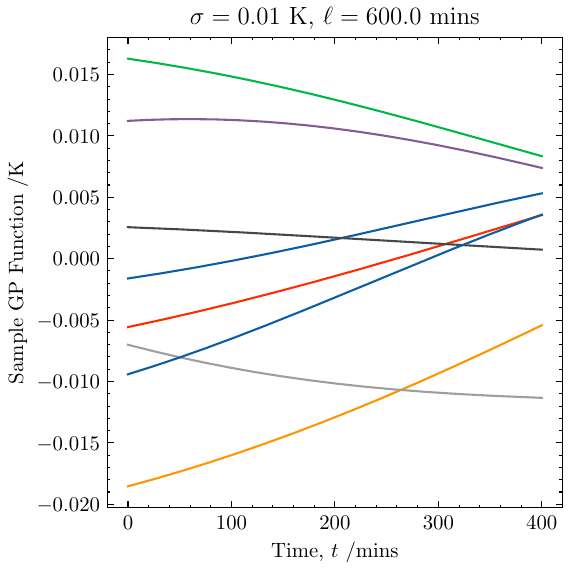}
    \caption{Example of eight functions drawn from the squared exponential Gaussian Process kernel given by equation \ref{e:SE_kernel}, with a scale factor of 0.01 K and a covariance length of 600 minutes.}
    \label{f:SE_kernel_samples}
\end{figure}

We choose this kernel as it has a simple form and describes a family of smooth functions, as seen in figure \ref{f:SE_kernel_samples}, of the form which we expect the systematic to take -- particularly for the smoothly modulated systematics simulated here. There are many other kernel choices which are valid, for example a periodic kernel \citep{rasmussenGaussianProcessesMachine2006} which will be useful which the systematic is modulated by the periodic galactic foreground power. We could also introduce a 2D Gaussian process kernel to incorporate correlations between frequency bins.

The covariance matrix we construct is hence
\begin{equation} \label{e:GP_covariance_matrix}
    \mathbf C_{ij} = K(t_i, t_j) = \sigma_{0,\text{GP}}^2 + K_\text{SE}(t_i,t_j),
\end{equation}
where $\sigma_{0,\text{GP}}$ is the Gaussian signal noise and is equivalent to adding a white noise kernel to the squared exponential kernel. We set the prior on $\sigma_{0,\text{GP}}$ to be a log uniform prior in the range $10^{-4} \leq \sigma_{0,\text{GP}} \leq 0.5$ K, on $\sigma_\text{SE}$ to be a log uniform prior in the range $0.01 \leq \sigma_\text{SE} \leq 0.5$ K and on $\ell$ to be a uniform prior in the range $100 \leq \ell \leq 1000$ minutes. The prior on the uncorrelated noise is taken from the standard REACH pipeline \citep{ansteyGeneralBayesianFramework2021a} while the scale factor and characteristic length priors are informed by the amplitude and time variance respectively of the foreground-modulated systematic we insert into the data. This covariance matrix is then combined with equation \ref{e:Gaussian_likelihood}, which will be referred to as the likelihood for the `Gaussian process pipeline', $\mathcal L_\text{GP}$.

Once the weighted mean hyperparameters, $\{\sigma_{0,\text{GP}}, \sigma_\text{SE}, \ell\}$, have been found using \textsc{PolyChord} and \textsc{Anesthetic} \citep{handleyAnestheticNestedSampling2019}, the mean regression line for a set of predicted times, $\mathbf t_\text{pred}$, using the observed data, $\{\mathbf t_\text{data}, \mathbf T_\text{data}\}$, can be found as
\begin{equation} \label{e:gp_regression_mean}
    \mu(\mathbf t_\text{pred}) = K(\mathbf t_\text{pred}, \mathbf t_\text{data}) K(\mathbf t_\text{data},\mathbf t_\text{data})^{-1} \mathbf T_\text{data},
\end{equation}
and the covariance matrix of the predicted data given by
\begin{align} \label{e:gp_regression_error}
\begin{split}
    \mathbf C_\text{pred} = &K(\mathbf t_\text{pred}, \mathbf t_\text{pred}) \\&- K(\mathbf t_\text{pred}, \mathbf t_\text{data}) K(\mathbf t_\text{data},\mathbf t_\text{data})^{-1} K(\mathbf t_\text{data},\mathbf t_\text{pred}).
\end{split}
\end{align}

\section{Results} \label{s:results}

In this section we will demonstrate the results of injecting a simulated sinusoidal systematic into both the standard and Gaussian process pipelines and compare the results of the two. In section \ref{s:signal_recovery} we will demonstrate the improvements to recovery of the 21cm signal made by the GP pipeline, in section \ref{s:parameter_sweep} we will see how systematics with different parameters affect the standard and GP pipelines, and in section \ref{s:regression} we will demonstrate a potential secondary use of the GP pipeline for regression of the time variation of the model residuals.

\subsection{Signal Recovery} \label{s:signal_recovery}

\begin{figure*}
     \centering
     \begin{subfigure}{0.49\textwidth}
         \centering
         \includegraphics[width=\textwidth]{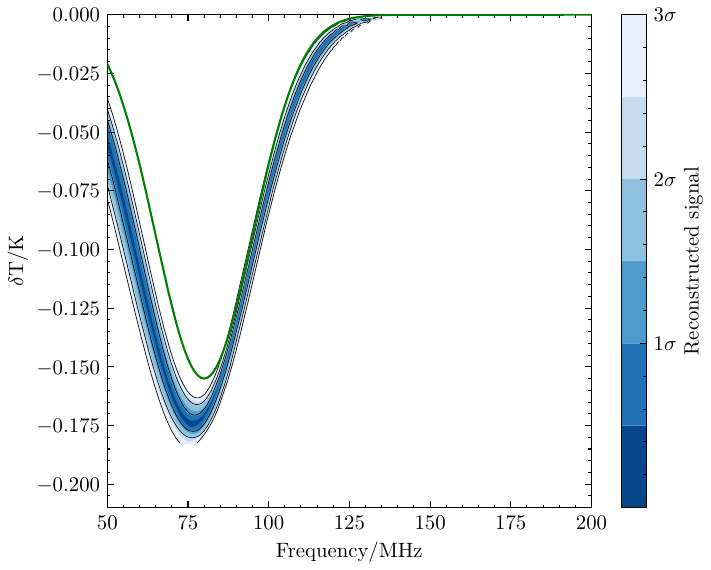}
         \caption{Standard Pipeline}
         \label{f:sig_comparison_std}
     \end{subfigure}
     \hfill
     \begin{subfigure}{0.49\textwidth}
         \centering
         \includegraphics[width=\textwidth]{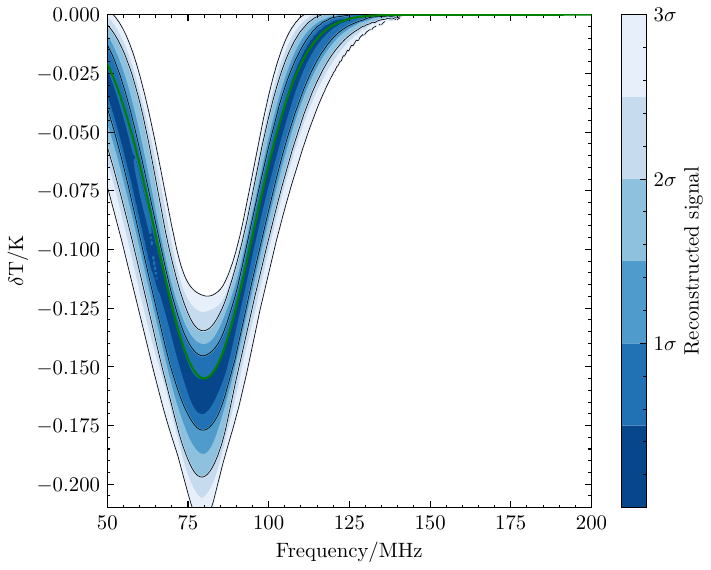}
         \caption{GP Pipeline}
         \label{f:sig_comparison_gp}
     \end{subfigure}
        \caption{Comparison of the recovered signal posteriors plotted with \textsc{fgivenx} \citep{handleyFgivenxPythonPackage2018} for the standard pipeline (equation \ref{e:standard_pipeline_likelihood}) and the Gaussian process pipeline (equations \ref{e:Gaussian_likelihood} and \ref{e:GP_covariance_matrix}). In green is plotted the `true' signal which was added into the simulated data. There is a foreground modulated systematic with an initial amplitude of $A_\text{sys} = A_{21} = 0.155$ K, period of $P_\text{sys} = 0.5 \sigma_{21} = 7.5$ MHz and phase of $\phi_\text{sys} = \pi$ present in the data.}
        \label{f:sig_comparison}
\end{figure*}

We first added a foreground modulated systematic over 24 time bins of length 15 minutes with an initial amplitude of $A_\text{sys} = A_{21} = 0.155$ K, period of $P_\text{sys} = 0.5 \sigma_{21} = 7.5$ MHz and phase of $\phi_\text{sys} = \pi$. For the chosen time period the foreground modulated systematic amplitude varies monotonically, increasing to a final amplitude of $\sim 1.7 A_\text{sys}$. The effects of adding a systematic to the data on the ability of the standard and GP pipelines to recover the signal can be seen in figure \ref{f:sig_comparison}. Plotted in green is the true signal and the blue contours show the 1, 2 and 3$\sigma$ contours of the signal posterior plotted with \textsc{fgivenx} \citep{handleyFgivenxPythonPackage2018}. It can be seen that while the standard pipeline misses the signal by greater than $3\sigma$, the GP pipeline has a much wider posterior, resulting in the pipeline capturing the signal within $1\sigma$. 

For the standard pipeline the mean noise parameter was $\sigma_{0,\text{std}} = 0.0637 \pm 0.0008$ K. In the case of the GP pipeline, the mean hyperparameters were $\sigma_{0,\text{GP}} = 0.0234 \pm 0.0003$ K, $\sigma_\text{SE} = 0.064 \pm 0.004$ K and $\ell = 640 \pm 60$ minutes. This shows the ability of the GP pipeline to separate the uncorrelated Gaussian noise from the correlated systematic, something which could not be done with a standard Gaussian likelihood which absorbs both noise and systematic in the $\sigma_{0,\text{std}}$ parameter.

When no systematic is added to the data both the standard and GP pipelines recovered the signal parameters to within $1\sigma$. For the standard pipeline the mean noise parameter was $\sigma_{0,\text{std}} = 0.0245 \pm 0.0003$ K. In the case of the GP pipeline, the mean hyperparameters were $\sigma_{0,\text{GP}} = 0.0245 \pm 0.0003$ K, $\sigma_\text{SE} = 0.0101 \pm 0.0009$ K and $\ell = 500 \pm 200$ minutes. In this case the posterior of the $\sigma_\text{SE}$ parameter has saturated at the lower end of its prior and can be assumed to either be very small or zero. Combining this with the fact that $\sigma_{0,\text{std}} = \sigma_{0,\text{GP}}$, this shows that the GP pipeline can be an important diagnostic tool to identify unmodelled systematics in the data, with the amplitude of the time-correlated noise parameter falling to zero in the absence of a systematic.

\subsection{Varying Systematic Parameters} \label{s:parameter_sweep}

To test the limits of the robustness of the standard and GP pipelines to systematics we varied the parameters of the simulated systematic to see how the goodness-of-fit and parameter biases changed. We repeated the pipeline runs with systematics with initial amplitudes of $A_\text{sys}/A_{21} = \{0.25, 0.50, 0.75, 1.00, 1.25, 1.50, 1.75, 2.00\}$, periods of $P_\text{sys}/ \sigma_{21} = \{0.5, 1.0, 1.5, 2.0, 2.5, 3.0, 3.5, 4.0\}$ and phases of $\phi_\text{sys} = \{0, 0.5 \pi, \pi, 1.5 \pi\}$. For each of these parameter combinations, both the standard and GP pipelines were run with time-separated data with 24 time bins of length 15 minutes -- equivalent to a single nights' observation.

We judge the goodness-of-fit of the pipeline fit using a root-mean-square error (RMSE) value. This is calculated as
\begin{equation} \label{e:rmse}
    \text{RMSE} = \left(\frac{\sum_i \sum_j w_j \big[T_\text{sg} (\nu_i, \theta^*_{\text{sg},j}) - T_\text{sg} (\nu_i, \theta_\text{true, sg})\big]^2}{N_\nu \sum_j w_j} \right)^\frac{1}{2},
\end{equation}
where $\theta^*_{\text{sg},j}$ and $w_j$ are the signal posterior samples and their weights respectively, $N_\nu$ is the number of frequency bins and $\theta_\text{true, sg}$ are the true signal parameters. 

\begin{figure*}
	\centering
	\includegraphics[width=0.9\linewidth]{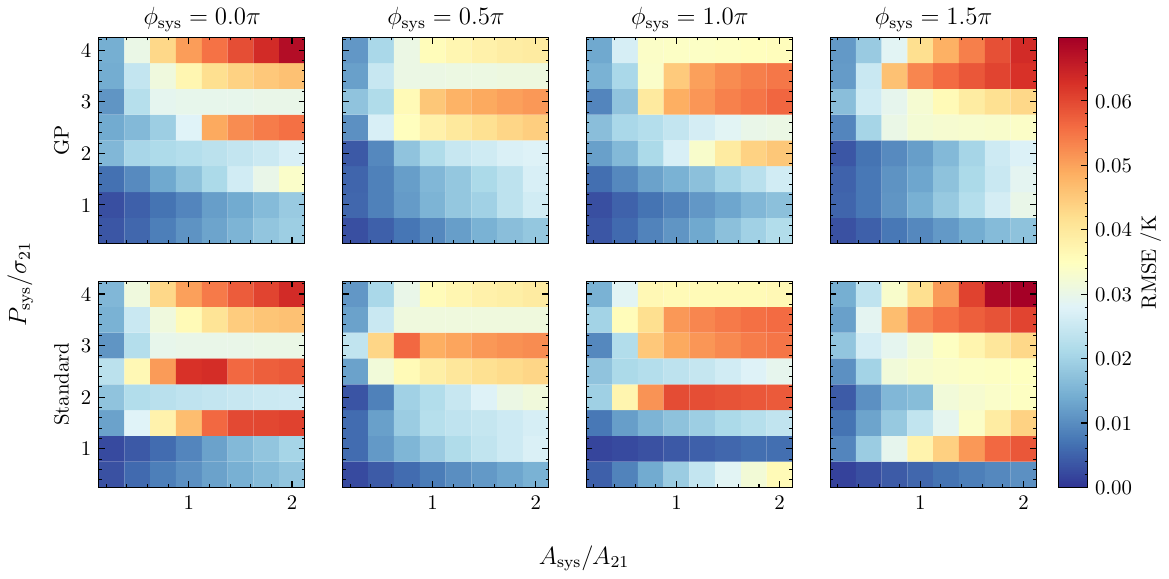}
	\caption{Weighted root mean square error (equation \ref{e:rmse}) of the pipeline fits for the standard pipeline (lower row) and the Gaussian processes pipeline (upper row) for different systematic parameters.}
	\label{f:rmse}
\end{figure*}

Figure \ref{f:rmse} shows the root mean square error for the standard pipeline (lower row) and the Gaussian process pipeline (upper row). It can be seen in most cases that the GP pipeline improves the goodness-of-fit (lower RMSE). This can be attributed to a combination of the GP pipeline widening the signal posterior whilst also reducing the biasing on the fitted signal parameters when a systematic is present in the data.

To compare how the confidence in the ability of the pipeline to detect the global signal, we use the Bayes factor
\begin{equation} \label{e:log_K}
    \log \mathcal K  = \log \mathcal{Z}_\text{GP} - \log \mathcal{Z}_\text{std},
\end{equation}
where $\mathcal{Z}_\text{GP}$ is the Bayesian evidence given when the GP pipeline is run, and $\mathcal{Z}_\text{std}$ is when the standard pipeline is run. This factor gives us the odds, $1 : \mathcal K$, that the data prefers GP pipeline over the standard pipeline \citep{mackayBayesianModelComparison1991}.

\begin{figure*}
	\centering
	\includegraphics[width=0.9\linewidth]{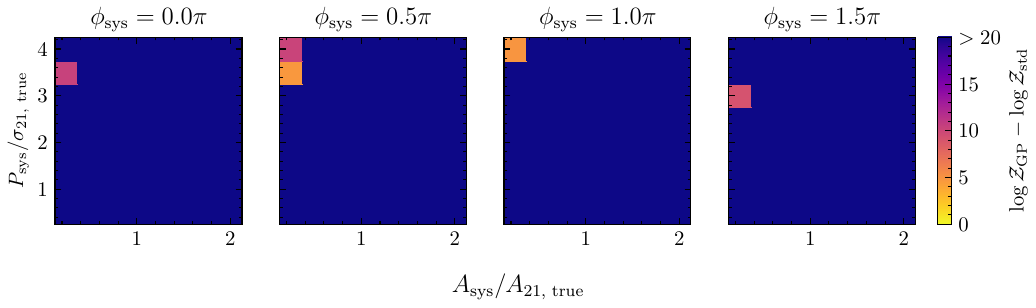}
	\caption{Log Bayes factor (equation \ref{e:log_K}) comparing the Bayesian evidence of the pipeline fits for the standard pipeline, $\mathcal Z_\text{std}$ and the Gaussian processes pipeline, $\mathcal Z_\text{GP}$, for different systematic parameters.}
	\label{f:log_K}
\end{figure*}

We see in figure \ref{f:log_K}, for all phases and most periods and amplitudes of the sinusoidal systematic, the Bayes factor is equal to or exceeds 20 indicating that the Gaussian process pipeline is highly favoured over the standard pipeline. In particular we find a minimum $\log \mathcal K$ value of 4.8, corresponding to minimum betting odds of around $1 : 120$ in favour of the GP pipeline's. We consider a Bayes factor over 2.5 to be a significant favouring and a Bayes factor over 5 to be a decisive favouring in line with the guidelines given by \cite{jeffreysTheoryProbability1998}. In the absence of a simulated systematic $\log \mathcal K = -64.0$, indicating a decisive preference for the standard pipeline over the GP pipeline -- a difference which may be attributed to the Occam penalty of the Bayesian evidence penalising the extra two GP hyperparameters \citep{hergtBayesianEvidenceTensortoscalar2021}.

\begin{figure*}
	\centering
	\includegraphics[width=0.9\linewidth]{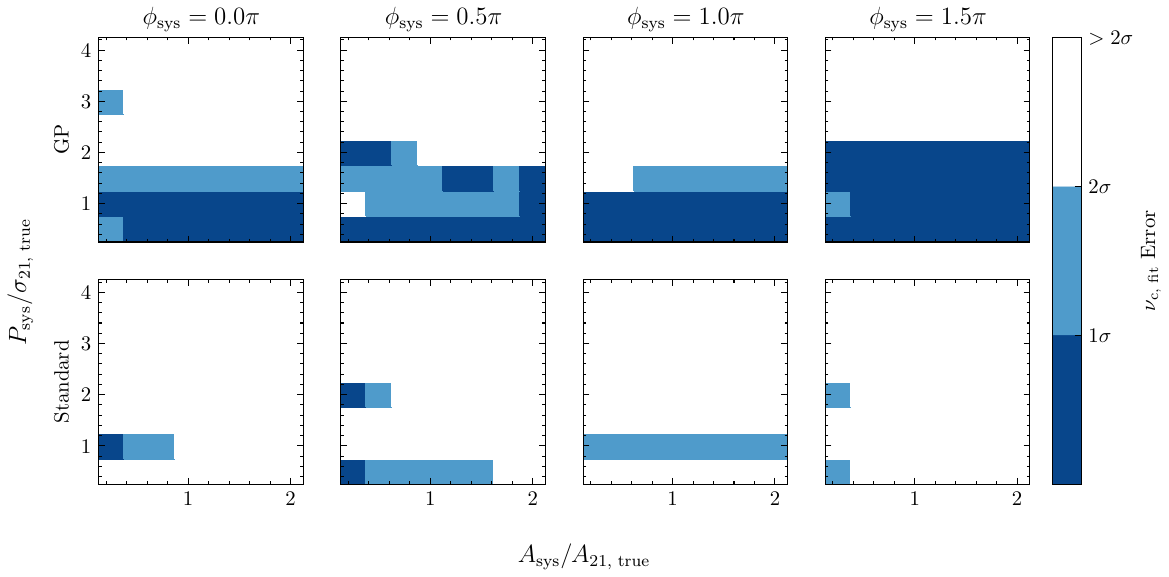}
	\caption{Error in the fitted value of the centre frequency of the Gaussian signal model for the standard pipeline (lower row) and the Gaussian processes pipeline (upper row) for different systematic parameters.}
	\label{f:nu_c}
\end{figure*}

\begin{figure*}
	\centering
	\includegraphics[width=0.9\linewidth]{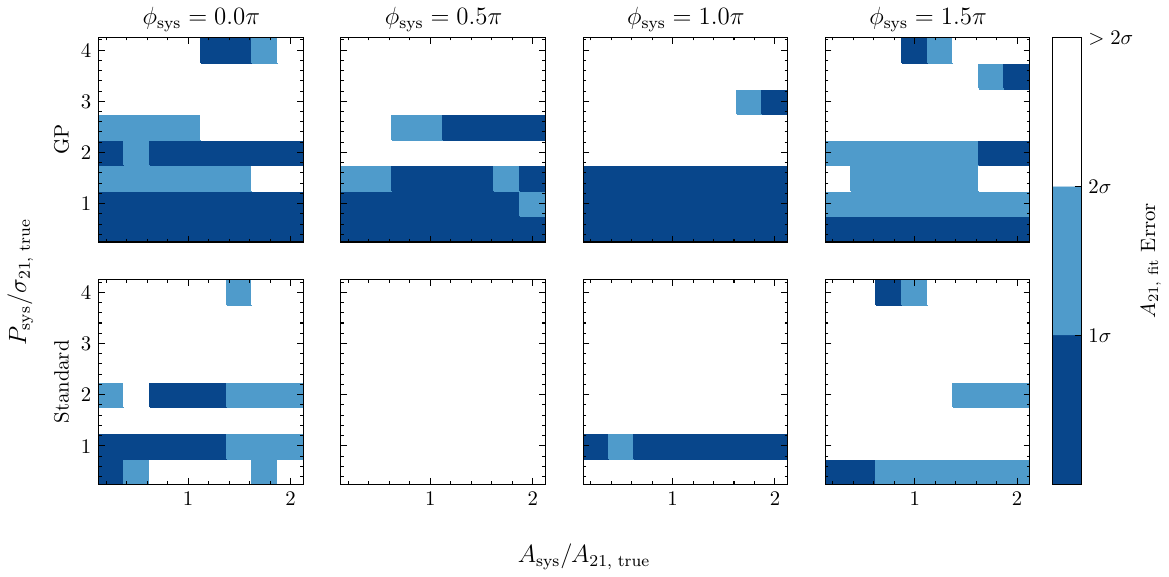}
	\caption{Error in the fitted value of the amplitude of the Gaussian signal model for the standard pipeline (lower row) and the Gaussian processes pipeline (upper row) for different systematic parameters.}
	\label{f:A_21}
\end{figure*}

\begin{figure*}
	\centering
	\includegraphics[width=0.9\linewidth]{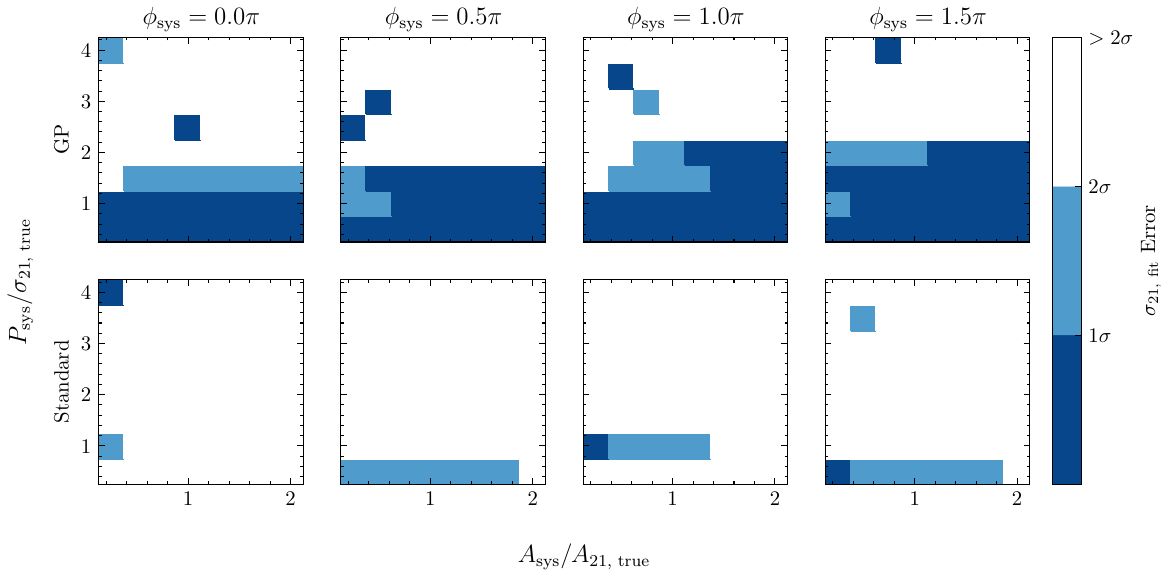}
	\caption{Error in the fitted value of the full-width half maximum of the Gaussian signal model for the standard pipeline (lower row) and the Gaussian processes pipeline (upper row) for different systematic parameters.}
	\label{f:width}
\end{figure*}

We now look at the individual signal parameters to more clearly see how the widening posterior alongside the reduction in bias affects the error in the parameter estimation. Figures \ref{f:nu_c}, \ref{f:A_21} and \ref{f:width} show the error in the recovery of the signal centre frequency, $\nu_c$, signal amplitude, $A_{21}$, and signal width $\sigma_{21}$, respectively for different systematics. It can be seen that there is a marked improvement in the recovery of the signal parameters when using the GP pipeline. In particular, when the systematic period is smaller than twice the signal width the GP pipeline is able to recover the parameters to within $2\sigma$ in most cases although has difficulty recovering $\nu_{21}$ when $\phi_\text{sys} = 0.5 \pi$ as the first trough of the systematic is close to the true signal centre frequency. For larger systematic periods there is a slight improvement in the RMSE but still miss the signal parameters by over 2$\sigma$, an effect which is likely caused by the troughs of the long-period sinusoids more closely representing the true Gaussian signal. In general varying the systematic amplitude has only a slight effect on the Gaussian process pipeline when the systematic period is low.

\begin{figure*}
	\centering
	\includegraphics[width=0.9\linewidth]{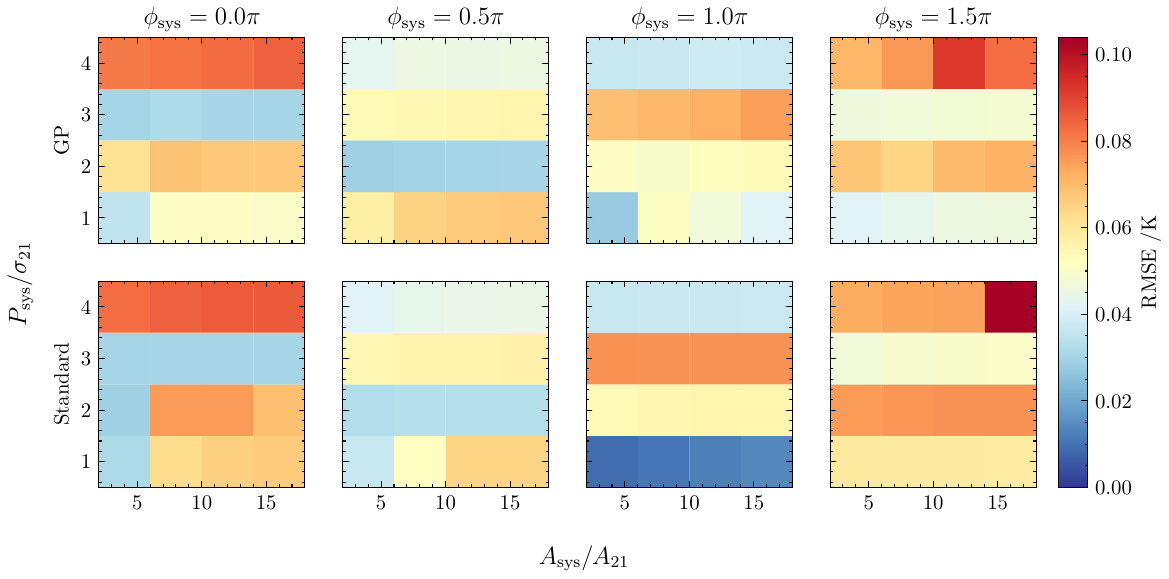}
	\caption{Weighted root mean square error (equation \ref{e:rmse}) of the pipeline fits for the standard pipeline (lower row) and the Gaussian processes pipeline (upper row) for systematics with amplitudes which greatly exceed that of the injected signal amplitude.}
	\label{f:rmse_lg_amp}
\end{figure*}

In order to discover the limits of the Gaussian process pipeline when run with large systematics, we also repeat the same parameter sweep with much greater systematic amplitudes of $A_\text{sys}/A_{21} = \{4.0, 8.0, 12.0, 16.0\}$ and periods of $P_\text{sys}/ \sigma_{21} = \{1.0, 2.0, 3.0, 4.0\}$. The RMSE values of these systematics are shown in figure \ref{f:rmse_lg_amp}. It can be seen that when the systematic amplitude four times the signal amplitude or greater, the GP pipeline no longer provides an improvement and in some cases worsens the RMSE value. Analysis of the posterior distributions of the hyperparameters shows that for the largest amplitude systematics the correlated noise hyperparameter, $\sigma_\text{SE}$, saturates at the higher end of its prior, demonstrating the need to adjust the original priors should there be a very large systematic in the data.

\begin{figure*}
	\centering
	\includegraphics[width=0.9\linewidth]{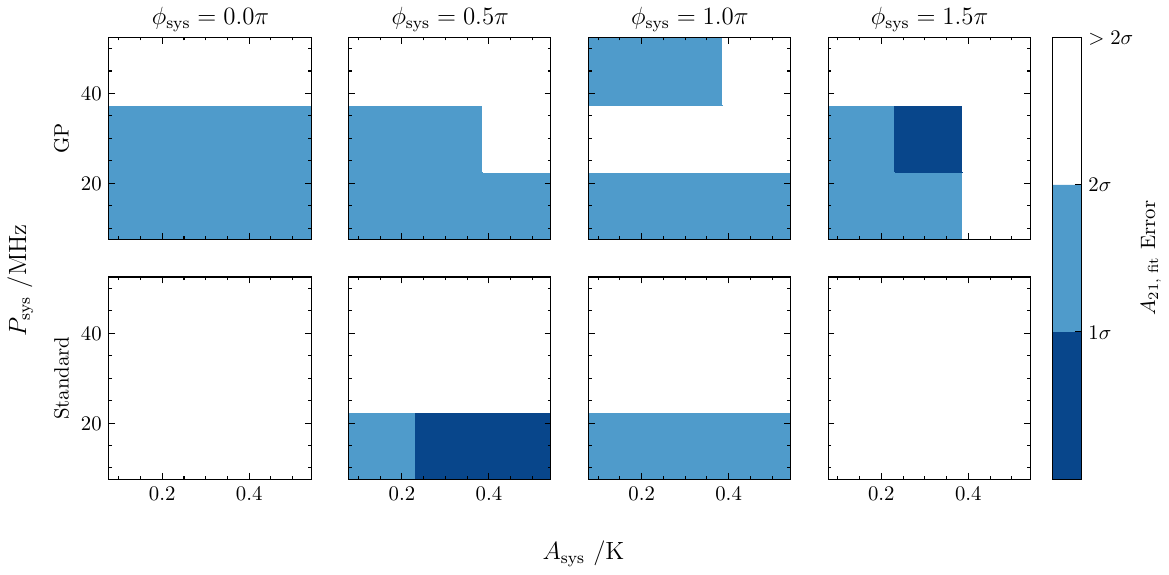}
	\caption{Distance from zero of the fitted value of the amplitude of the Gaussian signal model when there is no global signal in the data for the standard pipeline (lower row) and the Gaussian processes pipeline (upper row) for different systematic parameters.}
	\label{f:amp_no_sig}
\end{figure*}

Furthermore we can test the pipelines in the case where there is no global signal in the data but a Gaussian signal model is still being fitted for. Here we test the pipelines for systematics with amplitudes $A_\text{sys}/(0.155\text{ K}) = \{1.00, 2.00, 3.00\}$ and periods $P_\text{sys}/ (15\text{ MHz}) = \{1.0, 2.0, 3.0\}$. Figure \ref{f:amp_no_sig} shows the error in the fitted signal amplitude, where the true amplitude is 0 K. While the fit worsens somewhat at a phase of $0.5 \pi$ for low periods, in general there is an improvement in the ability to identify a lack of global signal in the data to within 2$\sigma$ by the Gaussian process pipeline. This shows that by taking into account the time correlation of the systematics, the Gaussian pipeline is less likely to fit a trough of the sinusoidal systematic.

\subsection{Gaussian Process Regression} \label{s:regression}

\begin{figure}
     \centering
     \includegraphics[width=\linewidth]{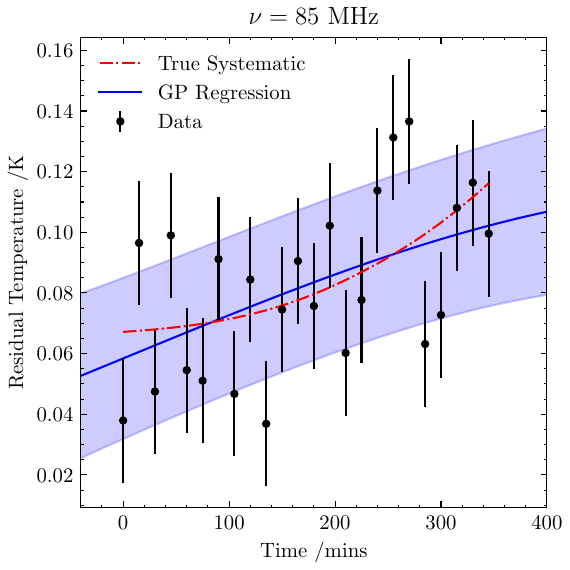}
    \caption{Example of Gaussian process time regression (equations \ref{e:gp_regression_mean} and \ref{e:gp_regression_error}) for the 85 MHz frequency bin for data with a systematic added. The blue line is the mean Gaussian process and the shaded area is the $\pm 1\sigma$ error. The red dot-dashed line is the true systematic amplitude over time. The initial amplitude of the systematic was set to $A_\text{sys} = 0.209$ K.}
    \label{f:gp_regression}
\end{figure}

Gaussian processes also enable investigation of systematic structure over time by allowing the calculation of a regression line for the model residuals. Using equations \ref{e:gp_regression_mean} and \ref{e:gp_regression_error} we get a mean temperature of the model residuals with time, indicating trends in the amplitude of the systematic.

Figure \ref{f:gp_regression} shows the Gaussian process time regression for the 85 MHz frequency bin for data with 24 time bins of length 15 minutes. A foreground-modulated systematic with an initial amplitude of $A_\text{sys} = 0.209$ K was added to the data. The black points are the residual temperatures after the mean foreground and signal models had been subtracted. The red dot-dashed line in the true signal amplitude and given by equation \ref{e:fg_modulated_systematic}. The blue line is the GP regression line determined using equation \ref{e:gp_regression_mean} and the blue shaded area is the $\pm 1\sigma$ error determined using equation \ref{e:gp_regression_error}.

While the regression line doesn't capture the same shape of the true systematic, the true line is within error of the regression and the GP captures the behaviour that the systematic is increasing with time. As the data is noisy it is unlikely that the GP will be able to capture the exact details of the systematic modulation. 

The method outlined in this paper is general and only assumes that the systematic varies with time, we can use the GP regression line to infer how we would expect the systematic to change with time. In future work where we may model and fit for time-varying systematics, continuing on from the time-averaged work of \cite{scheutwinkelBayesianEvidencedrivenDiagnosis2022a}, the GP regression is a useful base to inform the choice of model.

\section{Conclusions} \label{s:conclusions}

In this paper we presented a new method of mitigating the effects of unmodelled systematics when fitting for the global 21-cm signal in radio cosmology experiment data. By using a squared exponential Gaussian process kernel to fit for the correlations between time bins in the model residuals we are able to identify and mitigate the effects of time-varying residual systematics in the data.

We found that the Gaussian process pipeline was able to account for the presence of residual systematics in the data by widening the signal posteriors, reflecting our increased uncertainty in the signal parameters. We can use the squared-exponential kernel scale, $\sigma_\text{SE}$, to identify the presence of systematics in the data as its value will be non-zero should there be a systematic. This demonstrates the method's power as a diagnostic tool.

Furthermore we saw that the Gaussian process pipeline generally improved the goodness-of-fit, as measured using a root mean square value, with a 5\% improvement in RMSE values on average for the systematics we tested. Comparing the fitted signal parameters with the true values demonstrated that the GP pipeline reduces the biasing for systematics with a period less than twice the signal width. In many cases we found that the GP pipeline can recover parameters to within 1$\sigma$ despite the standard pipeline parameter fits being further than 2$\sigma$ from the true value. This can mainly be attributed to the widening of the posterior.

Further work will including using the Bayesian evidence to compare Gaussian Process kernels, as comparisons to other kernels could be beneficial to our understanding of the systematics. For example, a periodic kernel \citep{rasmussenGaussianProcessesMachine2006} could be useful when the systematic is modulated by the galactic foreground power as it is expected to vary periodically on a 24 hour timescale. A 2D Gaussian process could also be used to introduce correlations between frequency bins as well as the time bins.

\section*{Acknowledgements}

CJK would like to thank Erin Hayes and Will Handley for helpful discussion. CJK was supported by Science and Technology Facilities Council grant number ST/V506606/1. DJA was supported by Science and Technology Facilities Council grant number ST/X00239X/1. EdLA was supported by Science and Technology Facilities Council grant number ST/V004425/1. We would also like to thank the Kavli Foundation for their support of REACH.

\section*{Data Availability}

The data that supported the findings of this article will be shared on reasonable request to the corresponding author.



\bibliographystyle{mnras}
\bibliography{paper} 

\begin{thebibliography}{}
\makeatletter
\relax
\def\mn@urlcharsother{\let\do\@makeother \do\$\do\&\do\#\do\^\do\_\do\%\do\~}
\def\mn@doi{\begingroup\mn@urlcharsother \@ifnextchar [ {\mn@doi@}
  {\mn@doi@[]}}
\def\mn@doi@[#1]#2{\def\@tempa{#1}\ifx\@tempa\@empty \href
  {http://dx.doi.org/#2} {doi:#2}\else \href {http://dx.doi.org/#2} {#1}\fi
  \endgroup}
\def\mn@eprint#1#2{\mn@eprint@#1:#2::\@nil}
\def\mn@eprint@arXiv#1{\href {http://arxiv.org/abs/#1} {{\tt arXiv:#1}}}
\def\mn@eprint@dblp#1{\href {http://dblp.uni-trier.de/rec/bibtex/#1.xml}
  {dblp:#1}}
\def\mn@eprint@#1:#2:#3:#4\@nil{\def\@tempa {#1}\def\@tempb {#2}\def\@tempc
  {#3}\ifx \@tempc \@empty \let \@tempc \@tempb \let \@tempb \@tempa \fi \ifx
  \@tempb \@empty \def\@tempb {arXiv}\fi \@ifundefined
  {mn@eprint@\@tempb}{\@tempb:\@tempc}{\expandafter \expandafter \csname
  mn@eprint@\@tempb\endcsname \expandafter{\@tempc}}}

\bibitem[\protect\citeauthoryear{Anstey, {de~Lera~Acedo}  \& Handley}{Anstey
  et~al.}{2021}]{ansteyGeneralBayesianFramework2021a}
Anstey D.,  {de~Lera~Acedo} E.,   Handley W.,  2021, \mn@doi [Monthly Notices
  of the Royal Astronomical Society] {10.1093/mnras/stab1765}, 506, 2041

\bibitem[\protect\citeauthoryear{Anstey, {de Lera Acedo}  \& Handley}{Anstey
  et~al.}{2022}]{ansteyUseTimeDependent2022a}
Anstey D.,  {de Lera Acedo} E.,   Handley W.,  2022, Use of {{Time Dependent
  Data}} in {{Bayesian Global}} 21cm {{Foreground}} and {{Signal Modelling}}

\bibitem[\protect\citeauthoryear{Barkana}{Barkana}{2018}]{barkanaPossibleInteractionBaryons2018}
Barkana R.,  2018, \mn@doi [Nature] {10.1038/nature25791}, 555, 71

\bibitem[\protect\citeauthoryear{Bevins, Handley, Fialkov, {de Lera Acedo},
  Greenhill  \& Price}{Bevins et~al.}{2021}]{bevinsMAXSMOOTHRapidMaximally2021}
Bevins H. T.~J.,  Handley W.~J.,  Fialkov A.,  {de Lera Acedo} E.,  Greenhill
  L.~J.,   Price D.~C.,  2021, \mn@doi [Monthly Notices of the Royal
  Astronomical Society] {10.1093/mnras/stab152}, 502, 4405

\bibitem[\protect\citeauthoryear{Bevins, {de~Lera~Acedo}, Fialkov, Handley,
  Singh, Subrahmanyan  \& Barkana}{Bevins
  et~al.}{2022}]{bevinsComprehensiveBayesianReanalysis2022}
Bevins H. T.~J.,  {de~Lera~Acedo} E.,  Fialkov A.,  Handley W.~J.,  Singh S.,
  Subrahmanyan R.,   Barkana R.,  2022, \mn@doi [Monthly Notices of the Royal
  Astronomical Society] {10.1093/mnras/stac1158}, 513, 4507

\bibitem[\protect\citeauthoryear{Bolli, Bercigli, Ninni, Labate  \&
  Virone}{Bolli et~al.}{2020}]{bolliPreliminaryAnalysisEffects2020}
Bolli P.,  Bercigli M.,  Ninni P.~D.,  Labate M.~G.,   Virone G.,  2020, in
  2020 14th {{European Conference}} on {{Antennas}} and {{Propagation}}
  ({{EuCAP}}). pp~1--5, \mn@doi{10.23919/EuCAP48036.2020.9135350}

\bibitem[\protect\citeauthoryear{Bowman, Rogers  \& Hewitt}{Bowman
  et~al.}{2008}]{bowmanEmpiricalConstraintsGlobal2008}
Bowman J.~D.,  Rogers A. E.~E.,   Hewitt J.~N.,  2008, \mn@doi [The
  Astrophysical Journal] {10.1086/528675}, 676, 1

\bibitem[\protect\citeauthoryear{Bowman, Rogers, Monsalve, Mozdzen  \&
  Mahesh}{Bowman et~al.}{2018}]{bowmanAbsorptionProfileCentred2018}
Bowman J.~D.,  Rogers A. E.~E.,  Monsalve R.~A.,  Mozdzen T.~J.,   Mahesh N.,
  2018, \mn@doi [Nature] {10.1038/nature25792}, 555, 67

\bibitem[\protect\citeauthoryear{Cumner et~al.,}{Cumner
  et~al.}{2022}]{cumnerRadioAntennaDesign2022}
Cumner J.,  et~al., 2022, \mn@doi [J. Astron. Instrum.]
  {10.1142/S2251171722500015}, 11, 2250001

\bibitem[\protect\citeauthoryear{{De Oliveira-Costa}, Tegmark, Gaensler, Jonas,
  Landecker  \& Reich}{{De Oliveira-Costa}
  et~al.}{2008}]{deoliveira-costaModelDiffuseGalactic2008}
{De Oliveira-Costa} A.,  Tegmark M.,  Gaensler B.~M.,  Jonas J.,  Landecker
  T.~L.,   Reich P.,  2008, \mn@doi [Monthly Notices of the Royal Astronomical
  Society] {10.1111/j.1365-2966.2008.13376.x}, 388, 247

\bibitem[\protect\citeauthoryear{DeBoer et~al.,}{DeBoer
  et~al.}{2017}]{deboerHydrogenEpochReionization2017}
DeBoer D.~R.,  et~al., 2017, \mn@doi [PASP] {10.1088/1538-3873/129/974/045001},
  129, 045001

\bibitem[\protect\citeauthoryear{Dewdney, Hall, Schilizzi  \& Lazio}{Dewdney
  et~al.}{2009}]{dewdneySquareKilometreArray2009}
Dewdney P.~E.,  Hall P.~J.,  Schilizzi R.~T.,   Lazio T. J. L.~W.,  2009,
  \mn@doi [Proceedings of the IEEE] {10.1109/JPROC.2009.2021005}, 97, 1482

\bibitem[\protect\citeauthoryear{Dowell \& Taylor}{Dowell \&
  Taylor}{2018}]{dowellRadioBackground1002018}
Dowell J.,  Taylor G.~B.,  2018, \mn@doi [ApJL] {10.3847/2041-8213/aabf86},
  858, L9

\bibitem[\protect\citeauthoryear{Feng \& Holder}{Feng \&
  Holder}{2018}]{fengEnhancedGlobalSignal2018}
Feng C.,  Holder G.,  2018, \mn@doi [ApJL] {10.3847/2041-8213/aac0fe}, 858, L17

\bibitem[\protect\citeauthoryear{Field}{Field}{1959}]{fieldSpinTemperatureIntergalactic1959}
Field G.~B.,  1959, \mn@doi [The Astrophysical Journal] {10.1086/146653}, 129,
  536

\bibitem[\protect\citeauthoryear{Fixsen et~al.,}{Fixsen
  et~al.}{2011}]{fixsenARCADEMEASUREMENTABSOLUTE2011}
Fixsen D.~J.,  et~al., 2011, \mn@doi [ApJ] {10.1088/0004-637X/734/1/5}, 734, 5

\bibitem[\protect\citeauthoryear{Furlanetto, Peng~Oh  \& Briggs}{Furlanetto
  et~al.}{2006}]{furlanettoCosmologyLowFrequencies2006}
Furlanetto S.~R.,  Peng~Oh S.,   Briggs F.~H.,  2006, \mn@doi [Physics Reports]
  {10.1016/j.physrep.2006.08.002}, 433, 181

\bibitem[\protect\citeauthoryear{Handley}{Handley}{2018}]{handleyFgivenxPythonPackage2018}
Handley W.,  2018, \mn@doi [Journal of Open Source Software]
  {10.21105/joss.00849}, 3, 849

\bibitem[\protect\citeauthoryear{Handley}{Handley}{2019}]{handleyAnestheticNestedSampling2019}
Handley W.,  2019, \mn@doi [Journal of Open Source Software]
  {10.21105/joss.01414}, 4, 1414

\bibitem[\protect\citeauthoryear{Handley, Hobson  \& Lasenby}{Handley
  et~al.}{2015a}]{handleyPolychordNestedSampling2015}
Handley W.~J.,  Hobson M.~P.,   Lasenby A.~N.,  2015a, \mn@doi [Monthly Notices
  of the Royal Astronomical Society: Letters] {10.1093/mnrasl/slv047}, 450, L61

\bibitem[\protect\citeauthoryear{Handley, Hobson  \& Lasenby}{Handley
  et~al.}{2015b}]{handleyPolychordNextgenerationNested2015}
Handley W.~J.,  Hobson M.~P.,   Lasenby A.~N.,  2015b, \mn@doi [Monthly Notices
  of the Royal Astronomical Society] {10.1093/mnras/stv1911}, 453, 4384

\bibitem[\protect\citeauthoryear{Hergt, Handley, Hobson  \& Lasenby}{Hergt
  et~al.}{2021}]{hergtBayesianEvidenceTensortoscalar2021}
Hergt L.~T.,  Handley W.~J.,  Hobson M.~P.,   Lasenby A.~N.,  2021, \mn@doi
  [Phys. Rev. D] {10.1103/PhysRevD.103.123511}, 103, 123511

\bibitem[\protect\citeauthoryear{Hills, Kulkarni, Meerburg  \& Puchwein}{Hills
  et~al.}{2018}]{hillsConcernsModellingEDGES2018}
Hills R.,  Kulkarni G.,  Meerburg P.~D.,   Puchwein E.,  2018, \mn@doi [Nature]
  {10.1038/s41586-018-0796-5}, 564, E32

\bibitem[\protect\citeauthoryear{Jeffreys}{Jeffreys}{1998}]{jeffreysTheoryProbability1998}
Jeffreys S.~H.,  1998, The {{Theory}} of {{Probability}}, third edition, third
  edition edn.
Oxford {{Classic Texts}} in the {{Physical Sciences}}, {Oxford University
  Press}, {Oxford, New York}

\bibitem[\protect\citeauthoryear{Kraus, Tiuri, Raisanen  \& Carr}{Kraus
  et~al.}{1986}]{krausRadioAstronomyReceivers1986}
Kraus J.~D.,  Tiuri M.,  Raisanen A.~V.,   Carr T.~D.,  1986, Radio Astronomy
  Receivers.
{Cygnus-Quasar Books}

\bibitem[\protect\citeauthoryear{Leeney, Handley  \& Acedo}{Leeney
  et~al.}{2022}]{leeneyBayesianApproachRFI2022}
Leeney S. A.~K.,  Handley W.~J.,   Acedo E. d.~L.,  2022, A {{Bayesian}}
  Approach to {{RFI}} Mitigation (\mn@eprint {arxiv} {2211.15448}),
  \mn@doi{10.48550/arXiv.2211.15448}

\bibitem[\protect\citeauthoryear{MacKay}{MacKay}{1991}]{mackayBayesianModelComparison1991}
MacKay D.,  1991, in Advances in {{Neural Information Processing Systems}}.
  {Morgan-Kaufmann}

\bibitem[\protect\citeauthoryear{MacKay}{MacKay}{2002}]{mackayInformationTheoryInference2002}
MacKay D. J.~C.,  2002, Information {{Theory}}, {{Inference}} \& {{Learning
  Algorithms}}.
{Cambridge University Press}, {USA}

\bibitem[\protect\citeauthoryear{Monsalve et~al.,}{Monsalve
  et~al.}{2023}]{monsalveMapperIGMSpin2023}
Monsalve R.~A.,  et~al., 2023, Mapper of the {{IGM Spin Temperature}}
  ({{MIST}}): {{Instrument Overview}} (\mn@eprint {arxiv} {2309.02996}),
  \mn@doi{10.48550/arXiv.2309.02996}

\bibitem[\protect\citeauthoryear{Nambissan et~al.,}{Nambissan
  et~al.}{2021}]{nambissanSARASCDEoR2021}
Nambissan T.~J.,  et~al., 2021, \mn@doi [Exp Astron]
  {10.1007/s10686-020-09697-2}, 51, 193

\bibitem[\protect\citeauthoryear{Pattison, Anstey  \& {de Lera Acedo}}{Pattison
  et~al.}{2023}]{pattisonModellingHotHorizon2023}
Pattison J. H.~N.,  Anstey D.~J.,   {de Lera Acedo} E.,  2023, Modelling a
  {{Hot Horizon}} in {{Global}} 21 Cm {{Experimental Foregrounds}},
  \mn@doi{10.48550/arXiv.2307.02908}

\bibitem[\protect\citeauthoryear{Philip et~al.,}{Philip
  et~al.}{2019}]{philipProbingRadioIntensity2019}
Philip L.,  et~al., 2019, \mn@doi [J. Astron. Instrum.]
  {10.1142/S2251171719500041}, 08, 1950004

\bibitem[\protect\citeauthoryear{{Planck Collaboration} et~al.,}{{Planck
  Collaboration} et~al.}{2020}]{planckcollaborationPlanck2018Results2020a}
{Planck Collaboration} et~al., 2020, \mn@doi [Astronomy and Astrophysics]
  {10.1051/0004-6361/201833880}, 641, A1

\bibitem[\protect\citeauthoryear{Price et~al.,}{Price
  et~al.}{2018}]{priceDesignCharacterizationLargeaperture2018}
Price D.~C.,  et~al., 2018, \mn@doi [Monthly Notices of the Royal Astronomical
  Society] {10.1093/mnras/sty1244}, 478, 4193

\bibitem[\protect\citeauthoryear{Rasmussen}{Rasmussen}{2004}]{rasmussenGaussianProcessesMachine2004}
Rasmussen C.~E.,  2004, in Bousquet O.,  Von~Luxburg U.,   R{\"a}tsch G.,  eds,
  , Vol.~3176, Advanced {{Lectures}} on {{Machine Learning}}.
{Springer Berlin Heidelberg}, {Berlin, Heidelberg}, pp 63--71,
  \mn@doi{10.1007/978-3-540-28650-9_4}

\bibitem[\protect\citeauthoryear{Rasmussen \& Williams}{Rasmussen \&
  Williams}{2006}]{rasmussenGaussianProcessesMachine2006}
Rasmussen C.~E.,  Williams C. K.~I.,  2006, Gaussian Processes for Machine
  Learning.
Adaptive Computation and Machine Learning, {MIT Press}, {Cambridge, Mass}

\bibitem[\protect\citeauthoryear{{Razavi-Ghods} et~al.,}{{Razavi-Ghods}
  et~al.}{2023}]{razavi-ghodsReceiverDesignREACH2023}
{Razavi-Ghods} N.,  et~al., 2023, Receiver Design for the {{REACH}} Global
  21-Cm Signal Experiment, \mn@doi{10.48550/arXiv.2307.00099}

\bibitem[\protect\citeauthoryear{Reis, Fialkov  \& Barkana}{Reis
  et~al.}{2021}]{reisSubtletyLyPhotons2021a}
Reis I.,  Fialkov A.,   Barkana R.,  2021, \mn@doi [Monthly Notices of the
  Royal Astronomical Society] {10.1093/mnras/stab2089}, 506, 5479

\bibitem[\protect\citeauthoryear{Roberts, Osborne, Ebden, Reece, Gibson  \&
  Aigrain}{Roberts et~al.}{2013}]{robertsGaussianProcessesTimeseries2013}
Roberts S.,  Osborne M.,  Ebden M.,  Reece S.,  Gibson N.,   Aigrain S.,  2013,
  \mn@doi [Philosophical Transactions of the Royal Society A: Mathematical,
  Physical and Engineering Sciences] {10.1098/rsta.2011.0550}, 371, 20110550

\bibitem[\protect\citeauthoryear{Roque, Handley  \& {Razavi-Ghods}}{Roque
  et~al.}{2021}]{roqueBayesianNoiseWave2021}
Roque I. L.~V.,  Handley W.~J.,   {Razavi-Ghods} N.,  2021, \mn@doi [Monthly
  Notices of the Royal Astronomical Society] {10.1093/mnras/stab1453}, 505,
  2638

\bibitem[\protect\citeauthoryear{Scheutwinkel, {de Lera Acedo}  \&
  Handley}{Scheutwinkel
  et~al.}{2022}]{scheutwinkelBayesianEvidencedrivenDiagnosis2022a}
Scheutwinkel K.~H.,  {de Lera Acedo} E.,   Handley W.,  2022, \mn@doi [Publ.
  Astron. Soc. Aust.] {10.1017/pasa.2022.49}, 39, e052

\bibitem[\protect\citeauthoryear{Shaver, Windhorst, Madau  \& {de
  Bruyn}}{Shaver et~al.}{1999}]{shaverCanReionizationEpoch1999}
Shaver P.~A.,  Windhorst R.~A.,  Madau P.,   {de Bruyn} A.~G.,  1999, \mn@doi
  [Astronomy and Astrophysics] {10.48550/arXiv.astro-ph/9901320}, 345, 380

\bibitem[\protect\citeauthoryear{Sims \& Pober}{Sims \&
  Pober}{2020}]{simsTestingCalibrationSystematics2020}
Sims P.~H.,  Pober J.~C.,  2020, \mn@doi [Monthly Notices of the Royal
  Astronomical Society] {10.1093/mnras/stz3388}, 492, 22

\bibitem[\protect\citeauthoryear{Singh \& Subrahmanyan}{Singh \&
  Subrahmanyan}{2019}]{singhRedshifted21Cm2019}
Singh S.,  Subrahmanyan R.,  2019, \mn@doi [ApJ] {10.3847/1538-4357/ab2879},
  880, 26

\bibitem[\protect\citeauthoryear{Singh, Subrahmanyan, Shankar, Rao, Girish,
  Raghunathan, Somashekar  \& Srivani}{Singh
  et~al.}{2018}]{singhSARASSpectralRadiometer2018}
Singh S.,  Subrahmanyan R.,  Shankar N.~U.,  Rao M.~S.,  Girish B.~S.,
  Raghunathan A.,  Somashekar R.,   Srivani K.~S.,  2018, \mn@doi [Exp Astron]
  {10.1007/s10686-018-9584-3}, 45, 269

\bibitem[\protect\citeauthoryear{Singh et~al.,}{Singh
  et~al.}{2022}]{singhDetectionCosmicDawn2022}
Singh S.,  et~al., 2022, \mn@doi [Nat Astron] {10.1038/s41550-022-01610-5}, 6,
  607

\bibitem[\protect\citeauthoryear{Sivia}{Sivia}{2006}]{siviaDataAnalysisBayesian2006}
Sivia D.~S.,  2006, Data Analysis: A {{Bayesian}} Tutorial., 2nd ed. / d.s.
  sivia with j. skilling. edn.
Oxford Science Publications, {University Press}, {Oxford}

\bibitem[\protect\citeauthoryear{Tingay et~al.,}{Tingay
  et~al.}{2013}]{tingayMurchisonWidefieldArray2013}
Tingay S.~J.,  et~al., 2013, \mn@doi [Publications of the Astronomical Society
  of Australia] {10.1017/pasa.2012.007}, 30, e007

\bibitem[\protect\citeauthoryear{Wouthuysen}{Wouthuysen}{1952}]{wouthuysenExcitationMechanism21cm1952}
Wouthuysen S.~A.,  1952, \mn@doi [The Astronomical Journal] {10.1086/106661},
  57, 31

\bibitem[\protect\citeauthoryear{{de Lera Acedo} et~al.,}{{de Lera Acedo}
  et~al.}{2022}]{deleraacedoREACHRadiometerDetecting2022a}
{de Lera Acedo} E.,  et~al., 2022, \mn@doi [Nat Astron]
  {10.1038/s41550-022-01709-9}, 6, 984

\bibitem[\protect\citeauthoryear{van Haarlem et~al.,}{van Haarlem
  et~al.}{2013}]{haarlemLOFARLOwFrequencyARray2013}
van Haarlem M.~P.,  et~al., 2013, \mn@doi [A\&A] {10.1051/0004-6361/201220873},
  556, A2

\makeatother
\end{thebibliography}




\appendix


\bsp	
\label{lastpage}
\end{document}